\documentclass[12pt]{article}
\usepackage{amssymb}
\newtheorem{proposition}{Proposition}[section]  
\newcommand{\bprop}{\medskip\begin{proposition} ~~\\ \it}
\newcommand{\eprop}{\end{proposition} \hfill $\Box$ }
\newtheorem{naming}{Definition}[section]   
\newcommand{\bdefi}{\medskip\begin{naming} ~~\\ \it}
\newcommand{\edefi}{\end{naming} \hfill $\Diamond$ }
\newtheorem{example}{Example}[section]   
\def\bexam{\medskip\begin{example} ~~\\ \rm}
\def\eexam{\end{example} \hfill $\triangle$ }
\newcommand{\sect}[1]{\setcounter{equation}{0}\section{#1}}
\newcommand{\subsect}[1]{\subsection{#1}}
\newcommand{\subsubsect}[1]{\subsubsection{#1}}

\newcommand{\be}{\begin{equation}}
\newcommand{\ee}{\end{equation}}
\newcommand{\bea}{\begin{eqnarray}}
\newcommand{\eea}{\end{eqnarray}}
\newcommand{\bean}{\begin{eqnarray*}}
\newcommand{\eean}{\end{eqnarray*}}
\newcommand{\nn}{\nonumber}
\newcommand\IC{{\mathbb C}}

\newcommand\IM{{\mathbb M}}
\newcommand\IN{{\mathbb N}}

\newcommand\IZ{{\mathbb Z}}
\newcommand\IR{{\mathbb R}}

\def\abs#1{{\vert#1\vert}}

\def\inf1{{\cal L}^{(1,\infty)}}

\def\otr{\otimes_{\IR}}

\def\bar#1{\overline{#1}}

\def\bra#1{\left\langle #1\right|}
\def\ket#1{\left| #1\right\rangle}
\def\hs#1#2{\left\langle #1|#2\right\rangle}

\def\ca{{\cal A}}
\def\cb{{\cal B}}

\def\ce{{\cal E}}

\def\cu{{\cal U}}

\def\raw{\rightarrow}

\def\lrw{\leftrightarrow}
\begin{document}
\setcounter{page}{0}
\thispagestyle{empty}
\begin{flushright}
July 1999
\end{flushright}
\vspace{.5cm}
\begin{center}{\Large \bf Projective Modules of Finite Type \\ ~\\
over the Supersphere $S^{2,2}$}
\end{center} 
\vspace{1cm}
\centerline{\large  Giovanni Landi}
\vspace{5mm}
\begin{center}
{\it Dipartimento di Scienze Matematiche,
Universit\`a di Trieste \\ P.le Europa 1, I-34127, Trieste, Italy 
\\ and INFN, Sezione di Napoli, Napoli, Italy. \\
landi@mathsun1.univ.trieste.it}
\end{center}
\vspace{2.5cm}
\begin{abstract}
In the spirit of noncommutative geometry we construct all inequivalent
vector bundles over the $(2,2)$-dimensional supersphere $S^{2,2}$  
by means of global projectors $p$ via equivariant maps. Each
projector determines the projective module of finite type of sections of
the corresponding `rank $1$' supervector bundle over
$S^{2,2}$.  The canonical connection  $\nabla = p \circ d$ is used to
compute the Chern numbers by means of the Berezin integral
on $S^{2,2}$.
The associated connection $1$-forms are graded extensions of monopoles
with not trivial
topological charge.
Supertransposed projectors gives opposite values for the charges.
We also comment on the $K$-theory of  $S^{2,2}$.
\end{abstract}

\vfill
{\hfill \it This work is dedicated to Anna}

\newpage
\sect{Preliminaries and Introduction}
The Serre-Swan's theorem \cite{Sw,Co1} constructs a complete equivalence between the
category of (smooth) vector bundles over a (smooth) compact manifold $M$ and bundle
maps, and the category of finite projective modules over the commutative algebra $C(M)$
of (smooth) functions over $M$ and module morphisms. 
The space $\Gamma(M,E)$ of smooth sections of a vector bundle $E \raw M$ over a compact
manifold $M$ is a finite projective module over the commutative algebra $C(M)$ and
every finite projective $C(M)$-module can be realized as the module of sections of
some vector bundle over $M$. 

In the context of noncommutative geometry \cite{Co}, where a noncommutative algebra
${\cal A}$ is the analogue of the algebra of smooth functions on some `virtual
noncommutative space', finite projective (left/right) modules over ${\cal A}$ have been
used as algebraic substitutes for vector bundles, notably in order to construct
noncommutative gauge and gravity theories (see for instance, \cite{CFF,Co,CR,D-V,Mad}).
In fact, in noncommutative geometry there seems to be more that one
possibility for the analogue of the category of vector bundles \cite{D-V1}. 

On the other hand, there is a generalization of ordinary geometry which loosely
speaking goes under the name of {\it supergeometry}. Supergeometry can hardly be
considered noncommutative geometry and, indeed, one usually labels it {\it graded
commutative geometry}. In this paper we present a
finite-projective-module description
of all not trivial monopoles configurations on the $(2,2)$-dimensional
supersphere
$S^{2,2}$.  This will be done by constructing a suitable global projector  $p$ in the
graded matrix algebra
$\IM_{\abs{n},\abs{n} + 1}(G(S^{2,2}))$,
$n$ being the value of the topological charge, while $G(S^{2,2})$ denotes the graded
algebra of superfunctions on $S^{2,2}$. In the spirit of Serre-Swan theorem, the
projector $p$ determines the $G(S^{2,2})$-module $\ce$ of sections of the supervector
bundles on which monopoles live, as its image in the trivial module $G(S^2)^{2\abs{n} +
1}$ (corresponding to the trivial rank $(2\abs{n} + 1)$ supervector bundle over
$S^{2,2}$), i.e.  $\ce = p(G(S^2)^{2\abs{n} + 1})$. The value of the
topological charge is computed by taking the Berezin integral on $S^{2,2}$
of a suitable form. These monopoles will be also called Grassmann (or
graded) monopoles. 
A description of a Grassmann monopole on a supersphere, as a strong
connection in the framework of the theory of Hopf-Galois extensions, is in
\cite{DGH}.

We refer to \cite{book} for a friendly approach to modules of several kind
(including finite projective). Throughout the paper we shall avoid writing
explicitly the exterior product symbol for forms. 

\def\dia{^\diamond}
\def\edi{\eta\dia}
\sect{A Few Elements of Graded Algebra and Geometry}

For us, graded will be synonymous of $\IZ_2$-graded with the grading denoted as
follows. If $M = M_0 \oplus M_1$, then $\abs{m} = j$ means $m \in M_j$. The element $m$
is said to be homogeneous if either $m \in M_0$ or $m \in M_1$. Elements of $M_0$
(resp. of $M_1$) are called {\it even} (resp. {\it odd}). A morphism $\phi : M \raw N$ 
of graded structures is said to be {\it even} [ resp. {\it odd} ] if 
$\phi(M_j) \subseteq N_j$ [ resp. $\phi(M_j) \subseteq N_{j+1}$~, mod. $2$ ].

With $B_L = (B_L)_0 + (B_L)_1$ we shall indicate a real Grassmann algebra with $L$
generators. For simplicity we shall assume that $L < \infty$; mild assumptions 
(on the linear span of the products of odd elements) would
allow to treat the case $L = \infty$ as well. Here
$B_L$ is a graded commutative algebra, namely, 
\be
a b \in (B_L)_{\abs{a} + \abs{b}}~, ~~~ a b = (-1)^{\abs{a} \abs{b}} b a ~,
\ee
if the elements $a,b \in B_L$ are homogeneous. The algebra $B_L$ can also be written
as $B_L = \IR \oplus N_L$ with $N_L$ the nilpotent ideal. There are natural projections
$\sigma : B_L \raw \IR$ and $s : B_L \raw N_L$ which are called the {\it
body} and the {\it soul} maps respectively. The Cartesian product $B_L{}^{m+n}$ is made
into a graded $B_L$-module by setting
\bea
&& B_L{}^{m+n} = B_L{}^{m,n} ~\oplus~ B_L{}^{\bar{m},\bar{n}}~, \nn \\
&& B_L{}^{m,n} =: (B_L)_0{}^{m} \times (B_L)_1{}^{n}~, ~~~
   B_L{}^{\bar{m},\bar{n}} =: (B_L)_1{}^{m} \times (B_L)_0{}^{n}~.
\eea
The $(m,n)$-dimensional {\it superspace} $B_L{}^{m,n}$ is naturally a
$(B_L)_0$-module. If $L$ is finite, for consistency one must assume that $n \leq L$. A
body map $\sigma^{m,n} : B_L{}^{m,n} \raw \IR^{m}$ is defined by 
\be
\sigma^{m,n}(x^1, \dots, x^m ~;~ y^1, \dots, y^n) = \Big(\sigma(x^1), \dots,
\sigma(x^m)\Big)~.
\ee
This map is used to endow $B_L{}^{m,n}$ with a topology, called {\it De Witt topology},
whose open sets are the inverse images of open sets in $\IR^m$ through $\sigma^{m,n}$.

A graded $B_L$-module $M$ is said to be {\it free of dimension} $(m,n)$ if it is free
of rank $m+n$ over $B_L$ and has a basis formed by $m$ even and $n$ odd elements. The
module $M$ is said to be {\it projective (of finite type)} if it is the direct summand
of a free module (of finite dimensionality). 
Any right graded $B_L$-module $M$ can be turned into a left module, and viceversa, by
defining
\be
a m =: (-1)^{\abs{m}\abs{a}} m a~, ~~~\forall~ a\in B_L~, ~m\in M~,
\ee
which are homogeneous. Due to this fact, we shall not pay attention in  
distinguishing between right and left structures.

The collection $\IM_{m+n}(B_L)$ of $(m+n)\times(m+n)$ matrices with entry in $B_L$ is
a graded $B_L$-module of dimension $(m^2+n^2, 2mn)$. It is given a
grading in such a manner that its even part, denoted by $\IM_{m,n}(B_L)$,
is made of matrices of the form
\be\label{smatrix}
X = \left(
\begin{array}{cc}
A & B \\
C & D
\end{array}
\right)~.
\ee
Here $A$ and $D$ are $m\times m$ and $n\times n$ matrices respectively, both with
entries in $(B_L)_0$, whereas $B$ and $C$ are $m\times n$ and $n\times m$ matrices
respectively, both with entries in $(B_L)_1$. 
Odd matrices in $\IM_{m+n}(B_L)$ have the same form as (\ref{smatrix}) but
now  $A$ and $D$ have entries in $(B_L)_1$, whereas $B$ and $C$ 
have entries in $(B_L)_0$. \\
The $B_L$-module $\IM_{m+n}(B_L)$ is made a graded Lie $B_L$ algebra
(also, a Lie superalgebra over $B_L$) by endowing it with a graded bracket
whose definition on homogeneous elements is
\be
[X, Y] =: X Y - (-1)^{\abs{X} \abs{Y}} Y X~. 
\ee
The bracket is graded antisymmetric and satisfy a graded Jacobi identity. 
The {\it supertranspose} of the element (\ref{smatrix}) in 
$\IM_{m+n}(B_L)$ is defined as,
\be
\label{st}
\left(
\begin{array}{cc}
A & B \\
C & D
\end{array}
\right) ^{st} = 
\left(
\begin{array}{cc}
A^t & (-1)^{\abs{X}} ~C^t \\
- (-1)^{\abs{X}} ~B^t & D^t
\end{array}
\right) ~,
\ee
with the superscript $~^t~$ denoting usual matrix transposition. Then, one has
\be
(X Y)^{st} = (-1)^{\abs{X} \abs{Y}} ~Y^{st} X^{st}~.
\ee
In the present context, the ordinary trace $tr$ is replaced by the {\it supertrace}
$Str$ which, for a matrix of the form (\ref{smatrix}) is defined as
follows,
\be\label{strace}
Str(X) =: tr (A) - (-1)^{\abs{X}} tr (D)~.
\ee
The supertrace obeys graded versions of the usual properties of a trace. 
In particular, 
\bea
&& Str(X^{st}) = Str(X) \nn \\
&& Str(X Y) = (-1)^{\abs{X} \abs{Y}} Str(Y X) ~.
\eea

The collection $GL_{m,n}(B_L)$ of invertible matrices in $\IM_{m,n}(B_L)$
is naturally a super Lie group. If $H$ is any matrix in $GL_{m,n}(B_L)$,
one has that
\be
Str(H X H^{-1}) = Str(X)~.
\ee 
On elements of $GL_{m,n}(B_L)$ one defines a {\it 
superdeterminant} (or {\it Berezinian}) which is valued in $(B_L)_0^*$ ,
the group of invertible elements of $(B_L)_0$ \cite{BL}. First of all,
one proves that if the matrix $X$ has the form (\ref{smatrix}), then $X$ is
invertible if and only if $A$ and $D$ are invertible as ordinary matrices
with entries in $(B_L)_0$. Then, if $X \in GL_{m,n}(B_L)$, 
its superdeterminant is defined as 
\be\label{sdet}
Sdet(X) =: det(A - B D^{-1} C) ~det(D^{-1})~.
\ee
Again, the superdeterminant obeys graded versions of the usual properties of the
determinant \cite{BL,RS},
\bea
&& Sdet(X^{st}) = Sdet(X) \nn \\
&& Sdet(X Y) = (-1)^{\abs{X} \abs{Y}} Sdet(X) Sdet(Y)~,
\eea
for all $X,Y \in GL_{m,n}(B_L)$

All previous considerations and definitions concerning modules and matrices can be
extended to any graded commutative algebra. In particular, we shall consider graded
projective modules of finite type over graded commutative algebras of superfunctions
and matrices with entries in graded commutative algebras of superfunctions. 

In this paper we shall not dwell upon the different definitions of
superstructures while referring, for instance, to \cite{BBH}. Indeed, the only
supermanifolds we shall consider are the so called {\it De Witt supermanifolds}
\cite{DeW}. One says that a $(m,n)$-dimensional supermanifold $S$ is De Witt if it is
locally modeled on $B_L{}^{m,n}$ and has an atlas such that the images of the
coordinate maps are open in the De Witt topology of $B_L{}^{m,n}$. We shall denote by
$G^\infty(S, B_L)$ the graded $B_L$-algebra of $B_L$-valued supersmooth functions on the
supermanifold $S$. In a coordinate neighborhood, elements of $G^\infty(S, B_L)$ have a
usual superfield expansion in the odd coordinate functions.
Finally, we mention that it has been shown \cite{Ro} that a De Witt
$(m,n)$ supermanifold $S$ is a locally trivial fibre bundle over an ordinary
$m$-dimensional manifold $S_0$, with a vector fibre. The manifold $S_0$ is called the
body of $S$ and the bundle projection $\Phi : S \raw S_0$ is given in local bundle
coordinates by the body map $\sigma^{m,n}$.

Let $C_L =: B_L \otr \IC$ be the complexification of $B_L$. A {\it complex super line
bundle} over a supermanifold can be thought of either as a rank $(1,0)$ or a rank
$(0,1)$ super vector bundle since in both cases the standard fiber is $C_L$ while the
structure group is $(C_L)_0^* \simeq GL_{1,0}(C_L) \simeq GL_{0,1}(C_L)$, the group of
invertible even elements in
$C_L$. For this reason we shall not distinguish between these two cases and
refer to the final Section for additional remarks.  In the spirit of the
Serre-Swan theorem, supervector
bundles over De Witt supermanifolds will be `identified' with (finite) graded
projective modules of sections over the algebra of superfunctions over the base
supermanifold. This is due to the fact that, contrary to what happens for a general
supermanifold, any super vector bundle over a De Witt supermanifold admits a
connection \cite{BBH}. By the arguments in \cite{CQ} the existence of a
connection is
equivalent to the module of sections being projective.

Finally, we remind that, again in contrast with what happens for a general
supermanifold, complex super line bundles over a De Witt supermanifold are 
classified by their {\it obstruction class} and so they are in bijective
correspondence with elements in the integer sheaf cohomology group
$\check{H}^2(M,\IZ)$. For a line bundle, the obstruction class is
essentially the {\it first Chern class} of the bundle. By
using the morphism $j : \check{H}^2(M,\IZ) \raw H_{SDR}^2(M)$,
the latter being the de Rham cohomology group of superforms, the
obstruction class of complex super line bundles over a De Witt
supermanifolds $M$ can be realized as a super de Rham cohomology class of
$M$ \cite{BBH}. A representative for this class can be given in term of
the curvature of a connection on the bundle, the choice of the connection
being immaterial since different connections yield the same cohomology
class. We shall represent the Chern class of a complex line superbundle by
means of the curvature of a canonical connection which, is a sense, it is 
determined by the bundle itself.

\sect{The Hopf Fibration over the Supersphere $S^{2,2}$}

\subsect{The Supergroup $UOSP(1,2)$}
We shall describe the basic facts about the Lie supergroup  $UOSP(1,2)$ that we need
in this paper while referring to \cite{BT} for additional details.

Let $osp(1,2)$ be the Lie $B_L$ superalgebra of dimension $(3,2)$ with
even generators
${A_0, A_1, A_2}$ and odd generator ${R_+, R_-}$, explicitly given in matrix
representation by
\bea\label{osp12}
&&A_0 = {i \over 2}
\left(
\begin{array}{ccc}
0 & 0 & 0 \\
0 & 1 & 0 \\ 
0 & 0 & -1 
\end{array}
\right)~, ~~~
A_1 = {i \over 2}
\left(
\begin{array}{ccc}
0 & 0 & 0 \\
0 & 0 & 1 \\ 
0 & 1 & 0 
\end{array}
\right)~,
~~~
A_1 = {i \over 2}
\left(
\begin{array}{ccc}
0 & 0 & 0 \\
0 & 0 & -i \\ 
0 & i & 0 
\end{array}
\right)~, \nn \\
&&R_+ = {1 \over 2}
\left(
\begin{array}{ccc}
0 & -1 & 0 \\
0 & 0 & 0 \\ 
-1 & 0 & 0 
\end{array}
\right)~, ~~~
R_- = {1 \over 2}
\left(
\begin{array}{ccc}
0 & 0 & 1 \\
-1 & 0 & 0 \\ 
0 & 0 & 0 
\end{array}
\right)~.
\eea
Thus, a generic element $X\in osp(1,2)$ is written as $X = \sum_{k=0,1,2}
a_k A_k + \sum_{\alpha=+,-} \eta_\alpha R_\alpha$ with $a_k \in (B_L)_0,
~\eta_\alpha \in (B_L)_1$. The basis elements (\ref{osp12}) are closed under graded
commutator. In particular, the three even elements ${A_0, A_1, A_2}$ generate the Lie
algebra $so(2) \simeq su(2)$. 

If the integer $L$ is taken to be even, on the complexification $C_L = B_L \otr \IC$
there exists \cite{RS} an even graded involution
\bea
\dia &:& C_L ~\raw~ C_L~, \nn \\ 
&& \abs{x\dia} = \abs{x}~, ~~~\forall ~x\in
(C_L)_{\abs{x}}~, ~~~~~(c x)\dia = \bar{c} x\dia~, ~~~\forall ~c\in \IC~, ~x\in C_L ~,
\eea
which in addition verifies the properties 
\be(x y)\dia = x\dia ~y\dia~,  ~~~\forall ~x,y\in C_L~, ~~~~~
x {\dia}{\dia} = (-1)^{\abs{x}}x~, ~~~\forall ~x\in (C_L)_{\abs{x}}~.
\ee

The superalgebra $uosp(1,2)$ is defined to be the `real' subalgebra made of
elements of the form
\be
X = \sum_{k=0,1,2} a_k A_k + \eta R_+ +\edi R_-~, ~~~a_k \in (C_L)_0~, ~a_k\dia
= a_k~, ~~\eta\in (C_L)_1~.
\ee
Indeed, one introduces an adjoint operation $^\dagger$ which is defined on the bases
(\ref{osp12}) as
\be
A_i^\dagger = - A_i~, ~~i = 0,1,2~; ~~R_+^\dagger = - R_-~, 
~R_-^\dagger = R_+~,
\ee
and is extended to the whole of $C_L \otr osp(1,2)$ by using the involution $\dia$~.
Then, the superalgebra 
$uosp(1,2)$ is identified  as the collection of `anti-hermitian' elements
\be
uosp(1,2) = \{X \in C_L \otr osp(1,2) ~|~ X^\dagger = - X\}~.
\ee
The superalgebra $uosp(1,2)$ is the analogue of the compact real form of 
$C_L \otr osp(1,2)$. 

Finally, the Lie supergroup $UOSP(1,2)$ is defined to be the exponential map of
$uosp(1,2)$,
\be
UOSP(1,2) =: \{exp(X) ~|~ X \in uosp(1,2)\}~.
\ee
A generic element $s \in UOSP(1,2)$ can be presented as the product of
one-parameter subgroups,
\bea\label{s1ps}
&& s = u \xi ~, \nn \\
&& u = exp(a_0 A_0)exp(a_1 A_1)exp(a_1 A_1)~, ~~~a_k\dia = a_k \in (C_L)_0~, 
\nn \\
&& \xi = exp(\eta R_+)exp(\edi R_-) = exp(\eta R_+ +\edi R_-)~, ~~~\eta \in (C_L)_1~,
\eea
The last equality being a consequence of the nilpotency of the variable $\eta$.
Explicitly, the element
$s \in UOSP(1,2)$ can be parametrized as 
\be\label{supertotal}
s = \left(
\begin{array}{ccc}
1 + {1 \over 4}\eta \edi & - {1 \over 2}\eta & {1 \over 2}\edi \\
~&~&~\\
 -{1 \over 2}(a \edi - b\dia \eta) & ~a (1 - {1 \over 8}\eta \edi) ~
& ~-b\dia (1 - {1 \over 8}\eta \edi) ~ \\ 
 ~&~&~\\
~-{1 \over 2}(b \edi + a\dia \eta)~ & ~b (1 - {1 \over 8}\eta \edi) ~ 
& ~a\dia (1 - {1 \over 8}\eta \edi) ~
\end{array}
\right)~.
\ee
Here $a, b$ and $\eta$ are elements in the complex Grassmann algebra
$C_L$ with the restrictions $a_k\dia = a_k \in (C_L)_0$ and $\eta \in (C_L)_1$. 
Furthermore, the superdeterminant of the matrix (\ref{supertotal}) is
constrained to be $1$ and this yields the condition, 
\be\label{sdet1}
1 = Sdet(s) = a a\dia + b b\dia~.
\ee
It may be worth stressing that (\ref{sdet1}) is a condition in the even
part $(C_L)_0$ of the Grassmann algebra $C_L$ and thus involves all even
combinations of generators of the latter. \\
By using (\ref{s1ps}) one also finds the adjoint of any element to be
\be\label{supadj}
s^\dagger =: \xi^\dagger u^\dagger 
~=~ \left(
\begin{array}{ccc}
1 + {1 \over 4}\eta \edi & ~{1 \over 2}(a\dia \eta + b \edi)~  
& ~{1 \over 2}(b\dia \eta - a \edi)~ \\ 
~&~&~\\
{1 \over 2}\edi & ~a\dia (1 - {1 \over 8}\eta \edi)~
& ~b\dia (1 - {1 \over 8}\eta \edi)~ \\ 
~&~&~\\
 {1
\over 2}\eta & ~-b (1 - {1 \over 8}\eta \edi)~  & ~a(1 - {1 \over 8}\eta \edi)~
\end{array}
\right)~,
\ee
and checks that $s s^\dagger = s^\dagger s = 1$.

We shall also need $\cu(1)$, the Grassmann extension of
$U(1)$. It is realized as follows
\be
\cu(1) = \{ w \in (C_L)_{0} ~|~ w w\dia = 1 \}.
\ee
By embedding $\cu(1)$ in $UOSP(1,2)$ as 
\be
w ~\mapsto~ \left(
\begin{array}{ccc}
1 & 0 & 0 \\
0 & w & 0 \\ 
0 & 0 & w\dia~
\end{array}
\right)~,
\ee
we may think of $A_0$ as the generator of $\cu(1)$, i.e.
\be
\cu(1) \simeq \{ exp(\lambda A_0) ~|~ \lambda \in (C_L)_{0}~, ~\lambda\dia =
\lambda\}.
\ee

\subsect{The Principal $\cu(1)$ Bundle over $S^{2,2}$}

To our knowledge, the $\cu(1)$ principal fibration $\pi : UOSP(1,2) \raw
S^{2,2}$ over the $(2,2)$-dimensional supersphere was introduced for the 
first time in \cite{LM} and further studied in \cite{BBL} were, in
particular, it was shown that $S^{2,2}$ is a De Witt supermanifold over
the usual sphere $S^{2}$. The fibration is explicitly realized as follows. 
The total space is the $(1,2)$-dimensional supergroup $UOSP(1,2)$ while 
the structure supergroup is $\cu(1)$. 
We let $\cu(1)$ act on the right on $UOSP(1,2)$. If we parametrize any 
$s \in UOSP(1,2)$ by $s = s(a,b,\eta)$, then this action can be represented as
follows,
\be\label{cu1act}
UOSP(1,2) \times \cu(1) ~\raw~ UOSP(1,2)~, ~~~(s, w) \mapsto 
s \cdot w = s(aw,bw,\eta w)~. 
\ee
This action leaves unchanged the superdeterminant
\be
Sdet(s\cdot w) = aw (aw)\dia + bw (bw)\dia = a a\dia + b b\dia =1~.
\ee
The bundle projection
\bea
&& \pi : UOSP(1,2) ~\raw~ S^{2,2} =: UOSP(1,2) / \cu(1)~, \nn \\
&& \pi(a,b, \eta) =: (x_0, x_1, x_2, \xi_+, \xi_-)
\eea
can be given as the (co)-adjoint orbit through $A_0$ of the action of
$UOSP(1,2)$ on $uosp(1,2)$. With
$s^\dagger$ the adjoint of $s$ as given in (\ref{supadj}), one has
that 
\be
\pi(s) =: s ({2 \over i} A_0) s^\dagger =: \sum_{k=0,1,2} x_k ({2 \over i} A_k) +
\sum_{\alpha=+,-} \xi_\alpha (2 R_\alpha)~.
\ee
Explicitly, 
\bea\label{ss2coord}
&& x_0 = (a a\dia - b b\dia)(1 - {1 \over 4}\eta \edi) = 
(-1 + 2a a\dia)(1 - {1 \over 4}\eta \edi) = 
(1 - 2b b\dia)(1 - {1 \over 4}\eta \edi)~, \nn \\  
&& x_1 = (a\bar{b} + b\bar{a})(1 - {1 \over 4}\eta \edi)~, \nn \\ 
&& x_2 = i(a\bar{b} - b\bar{a})(1 - {1 \over 4}\eta \edi)~, \nn \\
&& \xi_- = -{1 \over 2}(a \edi + \eta b\dia) ~, \nn \\
&& \xi_+ = {1 \over 2}(\eta a\dia - b \edi)~.
\eea
One sees directly that the $x_k$'s are even, $x_k \in (C_L)_{0}$, and `real', $x_k
\dia = x_k$, while the $\xi_\alpha$ are odd, $\xi_\alpha \in (C_L)_{1}$, and such
that $\xi_-{\dia} = \xi_+$ (and $\xi_+{\dia} = -\xi_-$). In
addition, one finds that
\bea
\sum_{\mu=0}^2 (x_\mu)^2 + 2 \xi_- \xi_+ &=& 
(a a\dia + b b\dia)^2 (1 - {1 \over 2}\eta \edi) 
+ {1 \over 2}(a a\dia + b b\dia)\eta \edi \nn \\ 
&=& 1 ~.
\eea
Thus, the base space $S^{2,2}$ is a $(2,2)$-dimensional sphere in the superspace 
$B_L^{3,2}$. It turns out that $S^{2,2}$ is a De Witt supermanifold with {\it body}
the usual sphere $S^2$ in $\IR^3$ \cite{BBL}, a fact that we shall use later.
The inversion of (\ref{ss2coord}) gives the basic ($C_L$-valued) invariant
functions on $UOSP(1,2)$. Firstly, notice that 
\be
{1 \over 4}\eta \edi = \xi_-\xi_+~. 
\ee
Furthermore, 
\bea
&& a a\dia = {1 \over 2} \Big[1 +  x_0 (1 + \xi_-\xi_+)\Big] ~, \nn \\
&& b b\dia = {1 \over 2} \Big[1 -  x_0 (1 + \xi_-\xi_+)\Big] ~, \nn \\ 
&& a b\dia = {1 \over 2} (x_1 -i x_2)(1 + \xi_-\xi_+) ~, \nn \\
&& \eta a\dia = -(x_1 +i x_2)\xi_- + (1+x_0)\xi_+  ~, \nn \\
&& \eta b\dia = (x_1 -i x_2)\xi_+ - (1-x_0)\xi_-  ~,
\eea
a generic invariant (polynomial) function on $UOSP(1,2)$ being any function of the
previous variables. 

We shall denote with $\cb_{C_L} =: G^\infty(UOSP(1,2),C_L)$ the
graded algebra of $C_L$-valued smooth superfunctions on the total space $UOSP(1,2)$,
while
$\ca_{C_L} =: G^\infty(S^{2,2}, C_L)$ will be the graded algebra of $C_L$-valued smooth
superfunctions on the base space $S^{2,2}$. In the following, we shall identify
$\ca_{C_L}$ with its image in $\cb_{C_L}$ via pull-back.

\newpage

\sect{The Equivariant Maps and the Projectors}\label{se:emp}

\subsect{The Equivariant Maps}

Just as it happens for the group $U(1)$, irreducible representations of the
supergroup group
$\cu(1)$ are labeled by an integer
$n\in\IZ$, any two representations associated with different integers being
inequivalent. They can be explicitly given as left  representations on $C_L$, 
\be\label{repn}
\rho_n ~:~ \cu(1) \times C_L ~\raw~ C_L~, 
~~~(w,c) \mapsto \rho_n(w) \cdot c =: w^n c~.
\ee
In order to construct the corresponding equivariant maps $\varphi : UOSP(1,2) \raw C_L$
we shall distinguish between the two cases for which the integer $n$ is negative or
positive. From now on, we shall take the integer $n$ to be always positive 
and consider the two cases corresponding to $\mp n$.

\subsubsect{The Equivariant Maps for Negative Labels}

Given any positive integer, $n\in\IN$, 
the equivariant maps $\varphi_{-n} : UOSP(1,2) \raw C_L$ 
corresponding to the representation of $\cu(1)$ labelled by $-n$ 
are of the form \be\label{equi-n}
\varphi_{-n}(\eta, a,b) = {1 \over 2} ~\eta \sum_{j=0}^{n-1} a^{n-1-j} b^{j} f_j + 
(1 - {1 \over 8}\eta \edi) \sum_{k=0}^{n} a^{n-k} b^{k} g_k~,
\ee
with $f_j, ~j = 1, \dots, n-1$ and $g_k, ~k = 1, \dots, n$ any $C_L$-valued
functions which are invariant under the right action of $\cu(1)$ on $UOSP(1,2)$. The
reason for the choice of the additional invariant factor $(1 - {1 \over 8}\eta \edi)$
will be given later. Indeed,
\bea
\varphi_{-n}((\eta, a,b)w) &=& {1 \over 2} ~(\eta w) \sum_{j=0}^{n-1} (a w)^{n-1-j} (b
w)^{j} f_j +  (1 - {1 \over 8}\eta \edi) \sum_{k=0}^{n} (a w)^{n-k} (b w)^{k} g_k \nn
\\ &=& w^{n}\varphi_{-n}(\eta,a,b) \nn \\ 
&=& \rho_{-n}(w)^{-1} \varphi_{-n}(\eta, a,b)~.
\eea
The functions $f_j~, g_k$'s being invariant, we shall think of them as $C_L$-valued
functions on the base space $S^{2,2}$, namely as elements of the graded algebra
$\ca_{C_L}$. The space $G^\infty_{(-n)}(UOSP(1,2), C_L)$ of equivariant maps is a
right module over the (pull-back of) superfunctions
$\ca_{C_L}$.

\subsubsect{The Equivariant Maps for Positive Labels}

Given any positive integer, $n\in\IN$,
the equivariant maps $\varphi_{n} : UOSP(1,2) \raw C_L$
corresponding to the representation of $\cu(1)$ labelled by $n$
are of the form \be\label{equin}
\varphi_{n}(\eta, a,b) = {1 \over 2} ~\edi \sum_{j=0}^{n-1} (a\dia)^{n-1-j} (b\dia)^{j}
f_j +  (1 - {1 \over 8}\eta \edi) \sum_{k=0}^{n} (a\dia)^{n-k} (b\dia)^{k} g_k~,
\ee
with $f_j, ~j = 1, \dots, n-1$ and $g_k, ~k = 1, \dots, n$ any $C_L$-valued
functions which are invariant under the right action of $\cu(1)$ on $UOSP(1,2)$.
Indeed,
\bea
\varphi_{n}((\eta, a,b)w) &=& {1 \over 2} ~(\eta w)\dia \sum_{j=0}^{n-1} ((a
w)\dia)^{n-1-j} ((b w)\dia )^{j} f_j +  (1 - {1 \over 8}\eta \edi) \sum_{k=0}^{n}
(a\dia w)^{n-k} (b\dia w)^{k} g_k
\nn
\\ &=& (w\dia)^{n}\varphi_{-n}(\eta,a,b) \nn \\ 
&=& \rho_{n}(w)^{-1} \varphi_{n}(\eta, a,b)~,
\eea
where we have used the fact that $w\dia = w^{-1}$.
As before, we shall think of the functions $f_j~, ~g_k$'s as elements of the graded
algebra
$\ca_{C_L}$. And the space $G^\infty_{(n)}(UOSP(1,2), C_L)$ of equivariant maps will
again be a right module over $\ca_{C_L}$.

\subsect{The Projectors and the Bundles}

We are now ready to introduce the projectors. Again we shall take the integer $n$
to be positive and keep separated the two cases corresponding to $\mp n$.

\subsubsect{The Construction of the Projectors for Negative Labels}

Given any positive integer, $n \in \IN$,
let us consider the supervector-valued function with ($n, n+1$)-components 
given by, \bea\label{mon-n}
&& \bra{\psi_{-n}}=: 
\left( {1 \over 2 } \eta \left(a^n, \dots, 
~\sqrt{ 
{\scriptstyle        
 \addtolength{\arraycolsep}{-.5\arraycolsep}
 \renewcommand{\arraystretch}{0.5}
 \left( 
\begin{array}{c}
 \scriptstyle n - 1 \\
 \scriptstyle k  
\end{array} \scriptstyle \right)} 
}
a^{n-1-k} b^k~, \dots, b^{n-1} \right) ; \right. \nn \\
&& ~~~~~~~~~~~~~~~~~~~~~~~~~~~~~~ \left. (1 - {1 \over 8}\eta \edi) 
\left(a^n,
\dots,  ~\sqrt{ 
{\scriptstyle        
 \addtolength{\arraycolsep}{-.5\arraycolsep}
 \renewcommand{\arraystretch}{0.5}
 \left( 
\begin{array}{c}
 \scriptstyle n \\
 \scriptstyle k  
\end{array} \scriptstyle \right)} 
}
a^{n-k} b^k~, \dots, b^n \right) \right)~,
\eea
where 
${\scriptstyle        
 \addtolength{\arraycolsep}{-.5\arraycolsep}
 \renewcommand{\arraystretch}{0.5}
 \left( 
\begin{array}{c}
 \scriptstyle n \\
 \scriptstyle k  
\end{array} \scriptstyle \right)} 
= {n! \over{k! (n-k)!}}$ are the binomial coefficients. \\
The supervector-valued function (\ref{mon-n}) is normalized,
\bea
\hs{\psi_{-n}}{\psi_{-n}} &=& {1 \over 4}\eta \edi \sum_{j=0}^{n-1} 
{\scriptstyle        
 \addtolength{\arraycolsep}{-.5\arraycolsep}
 \renewcommand{\arraystretch}{0.5}
 \left( 
\begin{array}{c}
 \scriptstyle n - 1 \\
 \scriptstyle j  
\end{array} \scriptstyle \right)} 
a^{n-j-1} b^j (a\dia)^{n-j-1} (b\dia)^j \nn \\
&~& ~~~~~~~~~~ + (1 - {1 \over 4}\eta \edi) \sum_{k=0}^{n} 
{\scriptstyle        
 \addtolength{\arraycolsep}{-.5\arraycolsep}
 \renewcommand{\arraystretch}{0.5}
 \left( 
\begin{array}{c}
 \scriptstyle n \\
 \scriptstyle k  
\end{array} \scriptstyle \right)} 
a^{n-k} b^k (a\dia)^{n-k} (b\dia)^k \nn \\ 
&=& {1 \over 4}\eta \edi (a a\dia + b b\dia)^{n-1}+ (1 - {1 \over 4}\eta \edi) (a
a\dia + b b\dia)^n \nn \\ &=& 1~.
\eea
Then, we can construct a projector in $\IM_{n,n+1}(\ca_{C_L})$ by
\be\label{pro-n}
p_{-n} =: \ket{\psi_{-n}} \bra{\psi_{-n}}~.
\ee
It is clear that $p_{-n}$ is a projector,
\bea
&& p_{-n}^2 =: \ket{\psi_{-n}} \hs{\psi_{-n}}{\psi_{-n}} \bra{\psi_{-n}} 
= \ket{\psi_{-n}} \bra{\psi_{-n}} = p_{-n}~, \nn \\
&& p_{-n}^\dagger = p_{-n}~.
\eea
Moreover, it is of rank $1$ because its supertrace is the constant superfunction
$1$, \be
Str (p_{-n}) = \hs{\psi_{-n}}{\psi_{-n}} = 1~.
\ee
The $\cu(1)$-action will transform the vector (\ref{mon-n}) multiplicatively,
\be
\bra{\psi_{-n}} ~\mapsto~ \bra{(\psi_{-n})^w} = w^n \bra{\psi_{-n}}~, ~~~\forall ~w \in
\cu(1)~.
\ee
As a consequence the projector $p_{-n}$ is invariant,
\be
p_{-n} ~\mapsto~ (p_{-n})^w = \ket{(\psi_{-n})^w}\bra{(\psi_{-n})^w} =
\ket{\psi_{-n}}(w\dia)^n w^n \bra{\psi_{-n}} = \ket{\psi_{-n}}
\bra{\psi_{-n}} = p_{-n}
\ee 
(being $w\dia w = 1$), and its entries are functions on the base space $S^{2,2}$, that
is they are elements of
$\ca_{C_L}$ as it should be. Thus, the right module of sections $\Gamma^\infty(S^{2,2},
E^{(-n)})$ of the associated bundle is identified with the image of $p_{-n}$ in the
trivial rank
$(2n+1)$ module $(\ca_{C_L})^{2n+1}$ and the module isomorphism between sections and
equivariant maps is given by,
\bea\label{iso-n}
&& \Gamma^\infty(S^{2,2},E^{(-n)}) ~\lrw~ G^\infty_{(-n)}(UOSP(1,2), C_L)~, \nn \\
&& \sigma = p_{-n} 
\left(
\begin{array}{c}
f_0 \\ \vdots \\ g_n
\end{array}
\right) ~\lrw~ \varphi_\sigma (a,b) = 
\bra{\psi_{-n}} \left(
\begin{array}{c}
f_0 \\ \vdots \\ g_n
\end{array}
\right)  \nn \\ 
&& \varphi_\sigma (a,b) = 
{1 \over 2} ~\eta \sum_{j=0}^{n-1} 
\sqrt{ 
{\scriptstyle        
 \addtolength{\arraycolsep}{-.5\arraycolsep}
 \renewcommand{\arraystretch}{0.5}
 \left( 
\begin{array}{c}
 \scriptstyle n - 1\\
 \scriptstyle j 
\end{array} \scriptstyle \right)} 
}
a^{n-1-j} b^{j} f_j  + 
(1 - {1 \over 8}\eta \edi) \sum_{k=0}^{n} 
\sqrt{ 
{\scriptstyle        
 \addtolength{\arraycolsep}{-.5\arraycolsep}
 \renewcommand{\arraystretch}{0.5}
 \left( 
\begin{array}{l}
 \scriptstyle n \\
 \scriptstyle k  
\end{array} \scriptstyle \right)} 
}
a^{n-k} b^{k} g_k~, ~~~~~
\eea
with $f_j, ~j = 1, \dots, n-1$ and $g_k, ~k = 1, \dots, n$ generic elements in
$\ca_{C_L}$. By comparison with (\ref{equi-n}) it is obvious that the previous map is
a module isomorphism, the extra factors given by the binomial coefficients being
inessential to this purpose, since they could be absorbed in a redefinition of the
functions.

\subsubsect{The Construction of the Projectors for Positive Labels}

Given any positive integer, $n \in \IN$, let us 
consider the supervector-valued function with ($n, n+1$)-components 
given by, \bea\label{monn}
&& \bra{\psi_n}=: 
\left( {1 \over 2 } \edi \left((a\dia)^n, \dots, 
~\sqrt{ 
{\scriptstyle        
 \addtolength{\arraycolsep}{-.5\arraycolsep}
 \renewcommand{\arraystretch}{0.5}
 \left( 
\begin{array}{c}
 \scriptstyle n - 1 \\
 \scriptstyle k  
\end{array} \scriptstyle \right)} 
}
(a\dia)^{n-1-k} (b\dia)^k~, \dots, (b\dia)^{n-1} \right) ; \right. \nn \\
&& ~~~~~~~~~~~~~~~~~~ \left. (1 - {1 \over 8}\eta \edi) 
\left((a\dia)^n,
\dots,  ~\sqrt{ 
{\scriptstyle        
 \addtolength{\arraycolsep}{-.5\arraycolsep}
 \renewcommand{\arraystretch}{0.5}
 \left( 
\begin{array}{c}
 \scriptstyle n \\
 \scriptstyle k  
\end{array} \scriptstyle \right)} 
}
(a\dia)^{n-k} (b\dia)^k~, \dots, (b\dia)^n \right) \right)~.
\eea
The supervector-valued function (\ref{monn}) is
normalized,
\bea
\hs{\psi_n}{\psi_n} &=& -{1 \over 4}\edi \eta \sum_{j=0}^{n-1} 
{\scriptstyle        
 \addtolength{\arraycolsep}{-.5\arraycolsep}
 \renewcommand{\arraystretch}{0.5}
 \left( 
\begin{array}{c}
 \scriptstyle n - 1 \\
 \scriptstyle j  
\end{array} \scriptstyle \right)} 
(a\dia)^{n-j-1} (b\dia)^j a^{n-j-1} b^j 
\nn \\ &~& ~~~~~~~~~~ 
+ (1 - {1 \over 4}\eta \edi) \sum_{k=0}^{n} 
{\scriptstyle        
 \addtolength{\arraycolsep}{-.5\arraycolsep}
 \renewcommand{\arraystretch}{0.5}
 \left( 
\begin{array}{c}
 \scriptstyle n \\
 \scriptstyle k  
\end{array} \scriptstyle \right)} 
(a\dia)^{n-k} (b\dia)^k a^{n-k} b^k \nn \\ 
&=& {1 \over 4}\eta \edi (a a\dia + b b\dia)^{n-1}+ (1 - {1 \over 4}\eta \edi) (a
a\dia + b b\dia)^n \nn \\ &=& 1~.
\eea
Then, we can construct a projector in $\IM_{n,n+1}(\ca_{C_L})$ by
\be\label{pron}
p_n =: \ket{\psi_n} \bra{\psi_n}~.
\ee
It is clear that $p_n$ is a projector,
\bea
&& p_n^2 =: \ket{\psi_n} \hs{\psi_n}{\psi_n} \bra{\psi_n} 
= \ket{\psi_n} \bra{\psi_n} = p_n~, \nn \\
&& p_n^\dagger = p_n~.
\eea
Moreover, it is of rank $1$ because its supertrace is the constant superfunction
$1$, \be
Str (p_n) = \hs{\psi_n}{\psi_n} = 1~.
\ee
The $\cu(1)$-action will transform the vector (\ref{monn}) multiplicatively,
\be
\bra{\psi_n} ~\mapsto~ \bra{(\psi_n)^w} = (w\dia)^n \bra{\psi_n}~, ~~~\forall ~w \in
\cu(1)~.
\ee
As a consequence the projector $p_n$ is invariant,
\be
p_n ~\mapsto~ (p_n)^w = \ket{(\psi_n)^w}\bra{(\psi_n)^w} =
\ket{\psi_n}w^n (w\dia)^n\bra{\psi_n} = \ket{\psi_n}
\bra{\psi_n} = p_n~,
\ee 
and its entries are functions on the base space $S^{2,2}$, that
is they are elements of
$\ca_{C_L}$ as it should be. Thus, the right module of sections $\Gamma^\infty(S^{2,2},
E^{(n)})$ of the associated bundle is identified with the image of $p_n$ in the
trivial rank
$(2n+1)$ module $(\ca_{C_L})^{2n+1}$ and the module isomorphism between sections and
equivariant maps is given by,
\bea
&& \Gamma^\infty(S^{2,2},E^{(n)}) ~\lrw~ G^\infty_{(n)}(UOSP(1,2), C_L)~, \nn \\
&& \sigma = p_n 
\left(
\begin{array}{c}
f_0 \\ \vdots \\ g_n
\end{array}
\right) ~\lrw~ \varphi_\sigma (a,b) = 
\bra{\psi_n} \left(
\begin{array}{c}
f_0 \\ \vdots \\ g_n
\end{array}
\right)  \nn \\ 
&& \varphi_\sigma (a,b) = 
{1 \over 2} ~\edi \sum_{j=0}^{n-1} 
\sqrt{ 
{\scriptstyle        
 \addtolength{\arraycolsep}{-.5\arraycolsep}
 \renewcommand{\arraystretch}{0.5}
 \left( 
\begin{array}{c}
 \scriptstyle n - 1\\
 \scriptstyle j 
\end{array} \scriptstyle \right)} 
}
(a\dia)^{n-1-j} (b\dia)^{j} f_j \nn \\
&& ~~~~~~~~~~~~~~~~~~~~~~~~~~~~~~~~~~~+ 
(1 - {1 \over 8}\eta \edi) \sum_{k=0}^{n} 
\sqrt{ 
{\scriptstyle        
 \addtolength{\arraycolsep}{-.5\arraycolsep}
 \renewcommand{\arraystretch}{0.5}
 \left( 
\begin{array}{l}
 \scriptstyle n \\
 \scriptstyle k  
\end{array} \scriptstyle \right)} 
}
(a\dia)^{n-k} (b\dia)^{k} g_k~, 
\eea
with $f_j, ~j = 1, \dots, n-1$ and $g_k, ~k = 1, \dots, n$ generic elements in
$\ca_{C_L}$. By comparison with (\ref{equin}) it is obvious that the previous map is
a module isomorphism, the extra factors given by the binomial coefficients being
inessential to this purpose, since they could be absorbed in a redefinition of the
functions.

\bigskip
The vector superfunctions $\bra{\psi_{n}}$ and $\bra{\psi_{-n}}$ are one the
supertransposed of the other, that is, 
\be\label{bratra-+}
\bra{\psi_{n}} = (\ket{\psi_{-n}})^{st} = \bra{(\psi_{-n})^{\dia}}~, 
\ee
and the corresponding projectors are related by supertransposition,
\be
p_{n} = (p_{-n})^{st}~.
\ee
Thus, by transposing a projector we get an inequivalent one  (unless 
the projector is the identity).

\bigskip
\noindent {\bf Examples.} Here we give the explicit projectors corresponding to the
lowest values of the charges \cite{La}, 
\be\label{spromon}
p_{-1} = {1 \over 2} \left(
\begin{array}{l}
2 \xi_+ \xi_- ~;~~~ (x_1 + i x_2)\xi_- - (1 + x_0)\xi_+ ~;~~~
 -(x_1 - i x_2)\xi_+ + (1 - x_0)\xi_- \\ ~\\
-(x_1 - i x_2)\xi_+ - (1 + x_0)\xi_- ~;~~~ 1 + x_0 + \xi_+ \xi_- ~;~~~
 x_1 - i x_2 \\ ~\\ 
-(x_1 + i x_2)\xi_- - (1 - x_0)\xi_+ ~;~~~~~~ x_1 + i x_2 ~;~~~~~ 
1 - x_0 + \xi_+ \xi_-
\end{array}
\right), 
\ee

\be
p_{1} = {1 \over 2} \left(
\begin{array}{l}
2 \xi_+ \xi_- ~;~~~ -(x_1 - i x_2)\xi_+ - (1 + x_0)\xi_- ~;~~~
 -(x_1 + i x_2)\xi_- - (1 - x_0)\xi_+ \\ ~\\
-(x_1 + i x_2)\xi_- + (1 + x_0)\xi_+ ~;~~~ 1 + x_0 + \xi_+ \xi_- ~;~~~
 x_1 + i x_2 \\ ~\\ 
(x_1 - i x_2)\xi_+ - (1 - x_0)\xi_- ~;~~~~~~ x_1 - i x_2 ~;~~~~~ 
1 - x_0 + \xi_+ \xi_-
\end{array}
\right).
\ee

\noindent
It is evident that these projectors are one the supertransposed of the other.
They are both projectors in $\IM_{1,2}(\ca_{C_L})$.

\sect{The Connections and the Charges}\label{se:cc}

Associated with any projector there is a canonical connection. Let us first consider
the  projector $p_{-n}$. The connection is given by 
\bea\label{con-n}
&&\nabla_{-n} = p_{-n} \circ d ~:~ \Gamma^\infty(S^{2,2},E^{(-n)}) ~\raw~
\Gamma^\infty(S^{2,2},E^{(-n)}) \otimes_{\ca_{C_L}} \Omega^1(S^{2,2}, C_L), \nn \\
&&\nabla_{-n}(\sigma) =: \nabla_{-n} \Big(p_{-n}\ket{\ket{f}}\Big) =
p_{-n}\Big(\ket{\ket{df}} + dp_{-n}\ket{\ket{f}}\Big)~,
\eea
where we have used explicitly the identification $\Gamma^\infty(S^{2,2},E^{(-n)})
= p_{-n}(\ca_{C_L})^{2n+1}$. 

\noindent
The curvature $\nabla_{-n}^2 : \Gamma^\infty(S^{2,2},E^{(-n)}) \raw
\Gamma^\infty(S^{2,2},E^{(-n)}) \otimes_{\ca_{C_L}} \Omega^2(S^{2,2}, C_L)$ 
is found to be
\be\label{cur-n}
\nabla_{-n}^2 = p_{-n}(dp_{-n})^2 = \ket{\psi_{-n}}\hs{d \psi_{-n}}{d \psi_{-n}}
\bra{\psi_{-n}}~. 
\ee

\noindent
Then, by means of a matrix supertrace the first Chern class of
the superbundle determined by $p_{-n}$ is represented by the $2$-superform
\bea\label{smoncf}
C_1(p_{-n}) &=:& - {1 \over 2 \pi i} ~Str (p_{-n} (dp_{-n})^2) = - {1 \over 2 \pi i}
\hs{d
\psi_{-n}}{d \psi_{-n}} \nn \\ &=& - {1 \over 2 \pi i} 
\Big[(da da\dia + db db\dia)(n - {1 \over 4}\eta \edi) \nn \\
&& ~~~~~~~~~~~~~~~~~~~~~~~~ + {1\over 4}  (a da\dia + b db\dia)(\eta d\edi - \edi
d\eta) + {1\over 4} d\eta d\edi \Big]~, \nn \\
&=& - {1 \over 2 \pi i} \Big[n (da da\dia + db db\dia)
+{1\over 4}  d(a\edi) d(\eta a\dia) +
{1\over 4}  d(b\edi) d(\eta b\dia)\Big]~.
\eea
By using the coordinates on $S^{2,2}$ the previous $2$-superform results 
in
\bea
C_1(p_{-n}) &=& {n \over 4 \pi }(x_0 dx_1 dx_2 + x_1 dx_2 dx_0 + x_2 dx_0 dx_1)
(1 + 3 \xi_-\xi_+) \nn \\ 
&& ~~~~~ + {1 \over 4 \pi i}\Big[(dx_1 -i dx_2)\xi_+ d\xi_+ - (dx_1 +i dx_2)\xi_-
d\xi_- +  dx_0(\xi_- d\xi_+  + \xi_+ d\xi_-) \nn \\ 
&& ~~~~~~~~~~~~~~~~~ + (x_1 -i x_2)d\xi_+ d\xi_+ - (x_1 +i x_2)d\xi_-d\xi_-
-2x_0d\xi_-d\xi_+ \Big]~.
\eea

\noindent
Finally, to compute the corresponding first Chern number we need the Berezin
integral over the supermanifold $S^{2,2}$. This is a rather simple task due to the
fact that
$S^{2,2}$ is a De Witt supermanifold over the two-dimensional sphere $S^2$ in
$\IR^3$. Indeed, by using the natural morphism of forms ~$\widetilde{~} :
\Omega^2(S^{2,2}) \raw \Omega^2(S^2)$, the first Chern number yielded by
the superform $C_1(p_{-n})$ is computed as \cite{Br}
\be\label{smoncn}
c_1(p_{-n}) =: Ber_{S^{2,2}}\Big(C_1(p_{-n})\Big) =: 
\int_{S^2} \widetilde{C_1(p_{-n})}~.
\ee
It is straightforward to find the projected form $\widetilde{C_1(p_{-n})} \in
\Omega^2(S^2)$.  The bundle projection $\Phi: S^{2,2} \raw S^2$ on the body
manifold $S^2$ is explicitly realized in terms of the body map,
\be
\Phi(x_0,x_1,x_2;\xi_-,\xi_+) = (\sigma(x_0),\sigma(x_1),\sigma(x_2))~.
\ee
We recall that fermionic variables do not have body, thus they project into zero. On
the other side, by denoting  $\sigma_i = \sigma(x_i), ~i=0,1,2$, the $\sigma_i$'s are
cartesian coordinates for the sphere $S^2$ in $\IR^3$ and obey the condition
$(\sigma_0)^2 + (\sigma_1)^2 + (\sigma_2)^2 =1$. The projected form
$\widetilde{C_1(p_{-n})}$ is found to be
\be
\widetilde{C_1(p_{-n})} = {n \over 4\pi}(\sigma_0 d\sigma_1 d\sigma_2 + \sigma_1
d\sigma_2 d\sigma_0 + \sigma_2 d\sigma_0 d\sigma_1) = {n \over 4\pi} vol(S^2)~.
\ee
As a consequence
\be
c_1(p_{-n}) = Ber_{S^{2,2}}\Big(C_1(p_{-n})\Big) = {n \over 4 \pi} \int_{S^2} d
(vol(S^2)) = n~.
\ee

It is easy to check that the supertransposed projector $p_{n}$ is obtained from
$p_{-n}$ by exchanging 
$a \lrw a\dia, b \lrw b\dia$ and $\eta \raw -\edi, ~\edi \raw \eta$. This 
amounts to the exchange of coordinates $x_2 \raw -x_2$ and $\xi_- \raw -\xi_+,
~\xi_+ \raw \xi_-$~. It is than clear that the corresponding Chern class
is represented by
\be 
C_1(p_{n}) = - C_1(p_{-n})~, \label{smoncftra}
\ee
while the corresponding Chern number is given by,
\be
c_1(p_{n}) = -c_1(p_{-n}) = -n~. \label{smoncntra}
\ee
Having different topological charges the projectors $p_{-n}$ and $p_{n}$
and the corresponding super line bundles are clearly inequivalent. 

\section{Graded Monopoles of any Charge}

Now we are going to compute the connection $1$-form associated with the
canonical connection. We shall due this for the positive values of the
topological charge, the construction for the negative charges being the
same. Thus, given the connection
(\ref{con-n}), the corresponding connection $1$-form on the equivariant maps, 
\be
A_{-n} \in End_{\cb_{C_L}}\Big(G^\infty_{(-n)}(UOSP(1,2), C_L)\Big)
\otimes_{\cb_{C_L}} \Omega^1(UOSP(1,2), C_L)~,
\ee 
has a very simple expression in terms of 
the supervector-valued function $\ket{\psi_{-n}}$ \cite{La,La1}, 
\be\label{confor}
A_{-n} = \hs{\psi_{-n}}{d \psi_{-n}}~.
\ee
The associated covariant derivative on any $\varphi^\sigma \in
G^\infty_{(-n)}(UOSP(1,2), C_L)$ is given by
\be
\nabla_{-n} (\varphi^\sigma) =: \hs{\psi_{-n}}{\nabla_{-n}(\sigma)} 
= \Big(d + \hs{\psi_{-n}}{d \psi_{-n}}\Big) \varphi^\sigma~;
\ee 
here $\sigma \in \Gamma^\infty(S^{2,2},E^{(-n)})$ and we have used the
isomorphism (\ref{iso-n}). The connection form (\ref{confor}) is anti-hermitian, a
consequence of the normalization
$\hs{\psi_{-n}}{\psi_{-n}}=1$:
\be
(A_{-n})^\dagger =: \hs{d\psi_{-n}}{\psi_{-n}} = - \hs{\psi_{-n}}{d \psi_{-n}} =
-A_{-n}~.
\ee
Explicitly,
\be
A_{-n} = (n - {1 \over 4} \eta \edi)(a d a\dia + b d b\dia) + {1 \over 8}(\eta d \edi
+ \edi d \eta)~.
\ee
As for the connection $1$-form $A_{n}$ carrying a negative value of the
topological charge one finds
\be
A_{n} = -(n - {1 \over 4} \eta \edi)(a d a\dia + b d b\dia) 
- {1 \over 8}(\eta d \edi + \edi d \eta) = -A_{-n}~.
\ee

The connection form $A_{+1}$ which corresponds, we remind, to the values $-1$ for the
topological charge, was introduced for the first time in \cite{LM} and
extensively studied in \cite{BBL}.
Following what we did in the latter paper,
we name {\it Grassmann (or graded) monopole} of charge $\pm n$ the
connection $1$-form
$A_{\mp n}$. 

\sect{The $K$-theory of the Supersphere $S^{2,2}$}
Given a complex supervector bundle
$\pi : E \raw M$ over a supermanifold $M$ one can define even and odd 
Chern classes \cite{BBH}. Of
course, the classes of both type are even cohomology classes but they get an additional
graded label in $\IZ_2$. Thus, if the bundle is of rank $(r,s)$, so 
that it can be
though of as having typical fiber $C_L{}^{r,s}$ and structure supergroup
$GL_{r,s}(C_L)$, there are $r$ even Chern classes $C^{(0)}_j(E) \in
\check{H}^{2j}(M,\IZ)~, ~j = 1, \dots, r$~, and $s$ odd classes $C^{(1)}_k(E) \in
\check{H}^{2k}(M,\IZ)~,  ~j = 1, \dots, s$. Then, one proceeds to define even and odd
total Chern classes and even and odd Chern characters. The two kind of
classes come from the two possible projectivizations of the bundle
$E$, an even and odd projectivization respectively. 

If the bundle has rank $(1,0)$ there is only one even not trivial 
class $C^{(0)}_1(E)
\in \check{H}^{2}(M,\IZ)$ and no odd classes. If the bundle has rank 
$(0,1)$ there is 
only one odd  not trivial class $C^{(1)}_1(E) \in 
\check{H}^{2}(M,\IZ)$ and no even classes.  Both these two classes 
$C^{(0)}_1(E)$ and $C^{(1)}_1(E)$ can be realized as  
super de Rham cohomology classes of the base supermanifold  $M$ and, at least
when $M$ is a De Witt supermanifold, they can be given a representative in terms of the
curvature of a connection on the bundle, the choice of the particular connection being
immaterial up to cohomologous forms. 

Now, it should be clear that all analysis of this paper, especially the one in
Section~\ref{se:emp} and Section~\ref{se:cc}, can be carried over for bundles on
$S^{2,2}$ of both ranks $(1,0)$ and $(0,1)$. Then, from the constructions of this paper
we can conclude that the {\it reduced} $\widetilde{K}_0$ group of $S^{2,2}$ is the
graded additive supergroup made of two copies of $\IZ$,
\be
\widetilde{K}_0(S^{2,2}) = \IZ \oplus \IZ~,
\ee
the first copy being given an even degree while the second one gets an odd one. It is
somewhat suggestive to write
\be
\widetilde{K}_0(S^{2,2}) = \widetilde{K}_0(S^{2}) \oplus \Pi\widetilde{K}_0(S^{2}),
\ee
with $S^2$ the ordinary $2$-dimensional sphere and $\Pi$ denoting the parity
change functor \cite{Ma}; here $\widetilde{K}_0(S^{2})$ is thought of as a
$(1,0)$
supergroup so that $\Pi\widetilde{K}_0(S^{2})$ is a $(0,1)$ supergroup.

From what was said at the end of Section~\ref{se:cc} we know
that by taking the supertranspose of projectors there is a change in sign
in the  corresponding topological charge (Chern number). Thus,
supertransposing of projectors, although it is an isomorphism in `super'
$K$-theory is not the identity map.

\bigskip\bigskip\bigskip
 
\noindent
{\bf Acknowledgments}. 
I thank L. Dabrowski and P. Hajac for helpful discussions. 
I am grateful to U. Bruzzo for discussions and  for reading the 
compuscript. And I had helpful email and fax exchanges with K. Fujii. 

\bigskip\bigskip\bigskip
 
\bibliographystyle{unsrt}

\begin{thebibliography}{99}

\bibitem{BBH} C. Bartocci, U. Bruzzo, D. Hern\'andez Ruip\'erez, {\it The Geometry
of Supermanifolds} (Kluwer, 1991).

\bibitem{BBL} C. Bartocci, U. Bruzzo, G. Landi, {\it Chern-Simons Forms on Principal
Superfiber Bundles}, J. Math. Phys. {\bf 31} (1990) 45-54.

\bibitem{BL} F.A. Berezin, D.A. Leites, {\it Supermanifolds}, \\
Soviet Math. Dokl. {\bf 16} (1975) 1218-1222. \\
D.A. Leites, {\it Introduction to the theory of Supermanifolds}, \\
Russ. Math. Surv. {\bf 35} (1980) 1-64.

\bibitem{BT} F.A. Berezin, V.N. Tolstoy, {\it The Group with Grassmann
Structure $UOSP(1.2)$}, \\
Commun. Math. Phys. {\bf 78} (1981) 409-428.

\bibitem{Br} U. Bruzzo, {\it Berezin Integration on De Witt Supermanifolds}, \\
Isr. J. Math. {\bf 80} (1992) 161-169.

\bibitem{CFF} A.H. Chamseddine, G. Felder, J. Fr\"{o}hlich, {\it Gravity in
Non-Commutative Geometry}, Commun. Math. Phys. {\bf 155} (1993) 205-217. 

\bibitem{Co} A. Connes, {\it Noncommutative Geometry} (Academic Press, 1994).

\bibitem{Co1} A. Connes, {\it Non-commutative Geometry and Physics}, in: {\it
Gravitation and Quantization}, Les Houches, Session LVII (Elsevier Science B.V.,
1995).

\bibitem{CR} A. Connes, M. Rieffel, {\it Yang-Mills for Non-commutative Two-Tori},
in {\it Operator Algebras and Mathematical Physics}, Contemp. Math. {\bf 62}
(1987) 237-266. 

\bibitem{CQ} J. Cuntz, D. Quillen, {\it Algebra Extensions and
Nonsingularity}, \\
J. Amer. Math. Soc. {\bf 8} (1995) 251-289.
 
\bibitem{DGH} L. Dabrowski, H. Grosse, P.M. Hajac, 
{\it Strong Connections and Chern-Connes Pairing in the Hopf-Galois
Theory}, SISSA Project 84/99/FM, in preparation.

\bibitem{DeW} B. De Witt, {\it Supermanifolds}, 2nd edition (Cambridge
University Press, 1992).

\bibitem{D-V} M. Dubois-Violette, {\it D\'erivations et calcul diff\'erentiel
non commutatif}, \\
C. R. Acad. Sci. Paris, {\bf I 307} (1988) 403-408. 

\bibitem{D-V1} M. Dubois-Violette, {\it Some Aspects of Noncommutative
Differential Geometry}, \\
q-alg/9511027.

\bibitem{book} G. Landi, {\it An Introduction to Noncommutative Spaces and Their
Geometries} \\ (Springer, 1997).

\bibitem{La} G. Landi, {\it Deconstructing Monopoles and Instantons}, \\
e-Print Archive: math-ph/9812004; Rev. Math. Phys., in press.

\bibitem{La1} G. Landi, {\it Projective Modules of Finite Type and Monopoles over
$S^2$}, \\ 
e-Print Archive: math-ph/9905014.

\bibitem{LM} G. Landi, G. Marmo, {\it Extensions of Lie
Superalgebras and Supersymmetric Abelian Gauge Fields}, Phys. Lett. {\bf B193}
(1987) 61-66.

\bibitem{Mad} J. Madore, {\it An Introduction to Noncommutative
Differential Geometry and its Physical Applications}, LMS Lecture Notes 206,
1995; 2nd Edition 1999. 

\bibitem{Ma} Yu.I. Manin, {\it Gauge Field Theory and Complex Geometry},
(Springer, 1988).

\bibitem{RS} V. Rittenberg, M. Scheunert, {\it Elementary Construction of
Graded Lie Groups}, \\
J. Math. Phys. {\bf 19} (1978) 709-713.

\bibitem{Ro} A. Rogers, {\it A Global Theory of Supermanifolds}, J. Math. Phys. {\bf
21} (1980) 1352-1365.

\bibitem{Sw} R.G. Swan, {\it Vector Bundles and Projective Modules}, \\
Trans. Am. Math. Soc. {\bf 105} (1962) 264-277.

\end{thebibliography}

\end{document}